\documentclass{webofc}
\usepackage[varg]{txfonts}
\usepackage{natbib}
\usepackage{lineno}
\usepackage{hyperref}

\begin{document}

\title{Jet Single Shot Detection}

\author{
\firstname{Adrian Alan} \lastname{Pol}\inst{1}\thanks{\email{adrianalan.pol@cern.ch}} \and \firstname{Thea} \lastname{Aarrestad}\inst{1} \and \firstname{Katya} \lastname{Govorkova}\inst{1} \and \firstname{Roi} \lastname{Halily}\inst{4} \and \firstname{Tal} \lastname{Kopetz}\inst{4} \and \firstname{Anat} \lastname{Klempner}\inst{4} \and \firstname{Vladimir} \lastname{Loncar}\inst{1}\fnsep\inst{3} \and \firstname{Jennifer} \lastname{Ngadiuba}\inst{2} \and \firstname{Maurizio} \lastname{Pierini}\inst{1} \and \firstname{Olya} \lastname{Sirkin}\inst{4} \and \firstname{Sioni} \lastname{Summers}\inst{1}}

\institute{European Organization for Nuclear Research (CERN), Geneva, Switzerland \and California Institute of Technology, Pasadena, USA \and Institute of Physics Belgrade, Belgrade, Serbia \and CEVA, Herzliya, Israel}

\abstract{We apply object detection techniques based on Convolutional Neural Networks to jet reconstruction and identification at the CERN Large Hadron Collider. In particular, we focus on  CaloJet reconstruction, representing each event as an image composed of calorimeter cells and using a Single Shot Detection network, called Jet-SSD. The model performs simultaneous localization and classification and additional regression tasks to measure jet features. We investigate Ternary Weight Networks with weights constrained to \{-1, 0, 1\} times a layer- and channel-dependent scaling factors. We show that the quantized version of the network closely matches the performance of its full-precision equivalent.}

\maketitle

\section{Introduction}\label{SEC:Introduction}

The majority of particles produced at the CERN Large Hadron Collider (LHC) are unstable and immediately decay in different particles. When quarks and gluons are produced, QCD confinement prevents them from travelling across the detector. Instead, they shower other quarks and gluons, eventually hadronizing into particles. The result of this process is a {\em jet}, a collimated showers of particles with adjacent trajectories. Jets are key in many physics analyses done on the data collected by the LHC experiments,  e.g.~\cite{butterworth2008jet, skiba2007using, khachatryan2014search, aad2015search}. The procedure of classifying the origin of these jets, i.e. the nature of the particle that initiated the shower, known as {\em jet tagging}~\cite{adams2015towards, abdesselam2011boosted, altheimer2012jet, altheimer2014boosted} is a fundamental task for collision reconstruction at the LHC. Similarly, it is important to determine the jet energy, momentum, and mass.

Traditional approaches to jet tagging rely on features designed by experts that detect characteristic energy deposition patterns~\cite{plehn2010stop, larkoski2014soft, thaler2011identifying, larkoski2013energy, krohn2010jet, ellis2010recombination, dasgupta2013towards, dasgupta2013jet, dasgupta2015jet}. In recent years, several studies projected the lower level detector measurements of the emanating particles into an image, known as {\em jet images}. This opened the path to applying computer vision and machine learning techniques~\cite{cogan2015jet, pearkes2017jet, baldi2016jet, macaluso2018pulling, almeida2015playing, de2016jet, guest2016jet, barnard2017parton, butter2018deep, komiske2017deep, lin2018boosting, kasieczka2019machine, kasieczka2017deep}, with particular attention to Convolutional Neural Networks (CNNs)~\cite{lecun1998gradient}.

The goal of this paper is to extend this approach to the problem of jet clustering, e.g., to replace FastJet~\cite{cacciari2012fastjet} on computing architectures where parallel computing is more adequate. At the same time, we aim at demonstrating that jet clustering, mass measurement, and tagging could all be handled simultaneously. Besides the practical advantages of a single-shot approach to jet reconstruction, one would benefit from mutual learning when accomplishing more tasks at once. For instance, a classifier and a regression running at once can learn that calibration constants depend on the nature of the jet, an issue that is not handled with ad-hoc post-processing (see~\cite{Sirunyan:2019wwa} as an example).

With the luminosity increase expected in the future, traditional reconstruction algorithms might suffer from execution time scaling worse than linearly with the number of collisions happening in one bunch crossing. For this reason, it is worth investigating solutions that could execute many tasks at once, while retaining accuracy and benefiting from the additional speed up offered by parallel computing architectures. Deep neural networks, such as those used for computing vision tasks, are an obvious candidate.

On the other hand, memory consumption is also an important aspect to keep under control. To this purpose, we investigate the use of extreme quantization, up to ternary precision, which is applied already at training time to retain accuracy.

The remainder of this paper is structured as follows. In Sections~\ref{SEC:SSD} and \ref{SEC:Quant} we briefly review single-shot detection and efficient model design techniques. In Section~\ref{SEC:Dataset} we introduce the dataset and in Section~\ref{SEC:Model} model architecture, implementation details and training procedure. Finally, in Section~\ref{SEC:Results} we present the evaluation metric and results.

\section{Single-shot object detection}\label{SEC:SSD}

Object detection is a fundamental task in computer vision. It is defined as the classification of objects from predefined categories in the image along with their precise spatial locations. The spatial location and extent of an object can be defined coarsely using a bounding box, which is an axis-aligned rectangle tightly bounding the object. Instead, a precise pixel-wise segmentation mask corresponds to the segmentation task.

Starting from Overfeat Network~\cite{sermanet2013overfeat}, the field of object detection focused on using primarily CNNs as a building block, achieving state-of-the-art results in tasks such as face~\cite{zhang2016joint} or pedestrian detection~\cite{zhang2016faster}. For a general survey on this subject, see~\cite{zou2019object, liu2020deep}.

The deep learning-based object detection models are divided into two groups: one~\cite{redmon2017yolo9000, redmon2016you, fu2017dssd, zhou2019objects, lin2017focal} or two~\cite{girshick2014rich, ren2015faster, girshick2015fast, dai2016r, xu2018deep} stage detectors. Two-stage detectors tend to achieve better accuracy, while one-stage detectors are simpler and faster, hence more suitable to online tasks. 

\begin{figure}
  \centering
  \includegraphics[width=\linewidth]{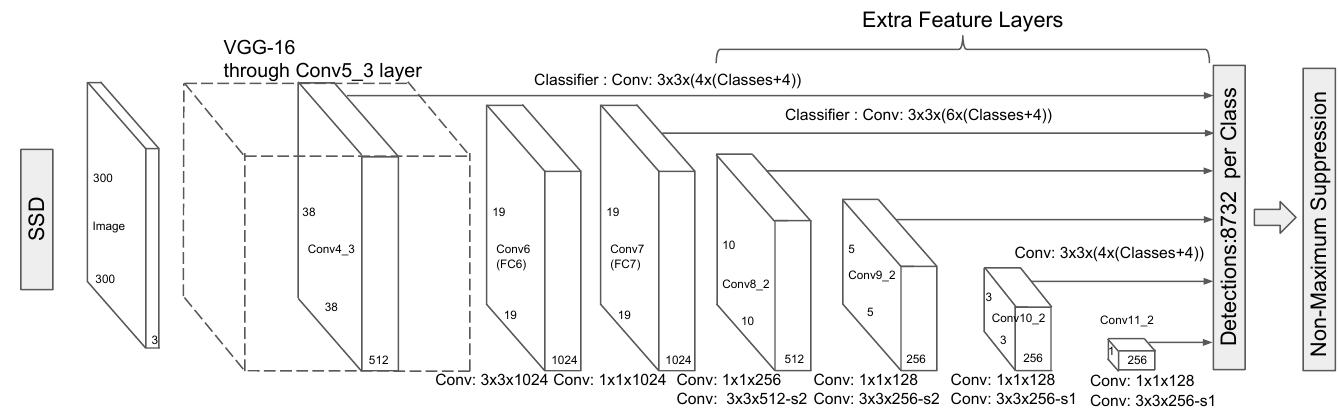}
  \caption{The architecture of the SSD network, proposed by~\cite{liu2016ssd}.}
  \label{fig:ssd}
\end{figure}

The Single Shot Mulibox Detector (SSD)~\cite{liu2016ssd}, shown in Figure~\ref{fig:ssd}, is a simple one-stage, anchor-based detector. First, a set of default regions in an image with a fixed shape and size is predefined to discretize the output space of bounding boxes, called anchors. These anchors have a diverse set of shapes to detect objects with different dimensions, i.e multiple scales and aspect ratios. At each location, the same amount of anchors is defined. Based on the ground truth, the object locations are matched with the most appropriate anchors to obtain the supervision signal for the anchor estimation.

During training, each anchor is refined by four box coordinates offsets (width, height, x and y) optimized by localization loss (a smooth L1 loss) and predict the categorical probabilities (including background), optimized by classification loss (categorical cross-entropy). To avoid a huge number of negative proposals dominating training gradients, hard negative mining is used to train the network, which fixes the foreground and background ratio.

The SSD architecture is fully convolutional, with initial layers based on a pre-trained backbone architecture, such as VGG-16~\cite{simonyan2014very}, followed by extra convolutional layers, progressively decreasing in size. The information in the last layer may be too coarse spatially to allow precise localization and at the same time, detecting large objects in shallow layers is non-optimal without large enough receptive fields. SSD performs detection over multiple scales by operating on multiple feature maps, i.e. at different depths of the network. Each of these feature maps is responsible for detecting objects according to their receptive field.

The final prediction is made by merging all detection results from different feature maps followed by a non-maximum suppression (NMS) step to produce the final detection. NMS removes duplicate predictions originating from multiple anchors.

\section{Efficient inference}\label{SEC:Quant}

Network compression~\cite{cheng2017survey} is a common technique to reduce the number of operations, model size, energy consumption, and over-training of deep neural networks. As neural network synapses and neurons can be redundant, compression techniques attempt to reduce the total number of them, effectively reducing multipliers. Several approaches have been successfully deployed without much loss in accuracy, including parameter pruning~\cite{lecun1989optimal, han2015deep, louizos2017learning} (selective removal of parameters based on a particular ranking and regularization), low-rank factorisation~\cite{sironi2014learning, denton2014exploiting, jaderberg2014speeding} (using matrix decomposition to estimate informative parameters), compact network architectures~\cite{szegedy2015going, howard2017mobilenets, iandola2016squeezenet, cohen2016group}, and knowledge distillation~\cite{bucilua2006model} (training a compact network with distilled knowledge of a large network).

A particularly successful compression technique is weight quantization~\cite{courbariaux2015binaryconnect, courbariaux2016binarized, zhou2016dorefa, rastegari2016xnor, hubara2017quantized, li2016ternary, zhu2016trained, lee2017lognet, cai2017deep}, which is reducing the precision of operations and operands. It has been observed that $32$-bit floating-point calculations or full-precision (FP) are not needed at inference to achieve optimal performance. Thus, reducing the precision of the calculations, i.e. weights and biases, has little impact on performance compared to speed up and resource usage. This includes moving away from floating point to fixed point, reducing bit-width and weight sharing. An example of a very aggressive strategy is reducing weight precision to ternary values restricted to $\{-1,0,1\}$ only, called Ternary Weight Network (TWN)~\cite{li2016ternary}. The quantization is performed during training, using a straight-through estimator~\cite{courbariaux2015binaryconnect}, where ternary weights are used during the forward and backward propagation but not during the parameters update. To make the network perform well, TWNs minimize the Euclidian distance between full precision weights and the ternary ones with the use of a non-negative layer- and channel-dependent scaling factor $\alpha$.

\section{Dataset}\label{SEC:Dataset}

The CERN LHC experiments implement a real-time selection process, called trigger~\cite{cms2016cms}, to store a fraction of the events for further analysis. Jets are useful for many measurements and physics searches. A truly minimal approach to perform identification and tagging is with jet images. Generally, jets need a component of tracks as well to be properly reconstructed. However, one could reconstruct the calorimeter part alone (known as CaloJet). The energy measurements of the emanating particles can be projected onto a cylindrical detector and represented as images by unfolding the inner surface of the calorimeter on a rectangle, and using the crystals as pixels, as in~\cite{, bhimji2018deep}.

The detector effects and hadronization have an important effect on the jet substructure. In this work, we use an emulation of the Compact Muon Solenoid (CMS) apparatus as a reference. There are two calorimeters within the solenoid volume of the CMS detector. A lead tungstate crystal Electromagnetic Calorimeter (ECAL) is designed to stop particles whose main interaction is electromagnetic (photons, electrons). A brass and scintillator Hadronic Calorimeter (HCAL) is designed to stop hadrons. They give a measurement of the energy of particles (charged and neutrals). Each of them is composed of a barrel and two endcap sections. Forward calorimeters extend the pseudorapidity range ($\eta$) coverage provided by the barrel ($\eta \leq 1.4$) and endcap detectors ($1.4 < |\eta| \leq 3.0$). A more detailed description of the CMS detector, together with a definition of the coordinate system used and the relevant kinematic variables, can be found in~\cite{collaboration2008cms}.

This study aims at identifying different kinds of jets. To this purpose, we consider $13$~TeV proton-proton collision events, in which RS gravitons decay to $b \bar b$, HH, WW, ZZ, or $t \bar t$ final states. Events are generated with Pythia~\cite{sjostrand2008brief} and the CMS detector effects are emulated using the Delphes~\cite{de2014delphes} library. In addition to the hard collision, parasitic {\it pileup} collisions are also simulated, overlapping minimum bias events. The number of pileup collisions is sampled from a Poisson distribution. The calorimeter cells (towers) in the barrel region are arranged in a fixed discrete space with fine segmentation in $\eta$, $\phi$, where $\phi$ is the translated azimuthal angle. The final image is formed by translating the calorimeter energy deposits into pixels, which results in a $340\times360$ pixel image. The intensity of each pixel is proportional to the sum of the energy of the corresponding cell. The previous studies on jet images implemented data pre-processing steps such as translation, rotation, re-pixelation, or inversion. However, in our study we only limit the input to barrel and endcap section, $\eta \in (-3, 3)$, and normalize pixel intensities to a fixed range <$0, 1$>, using maximum scaling. The ground truth labels for jets above threshold momentum ($30$~GeV/c for b and $200$~GeV/c for the jets from boosted heavy particles) are obtained using a simple cone algorithm, i.e. associating together particles whose trajectories lie within a circle of radius $R=0.4$ from the jet centre.

\begin{figure}
  \centering
  \includegraphics[width=.49\linewidth]{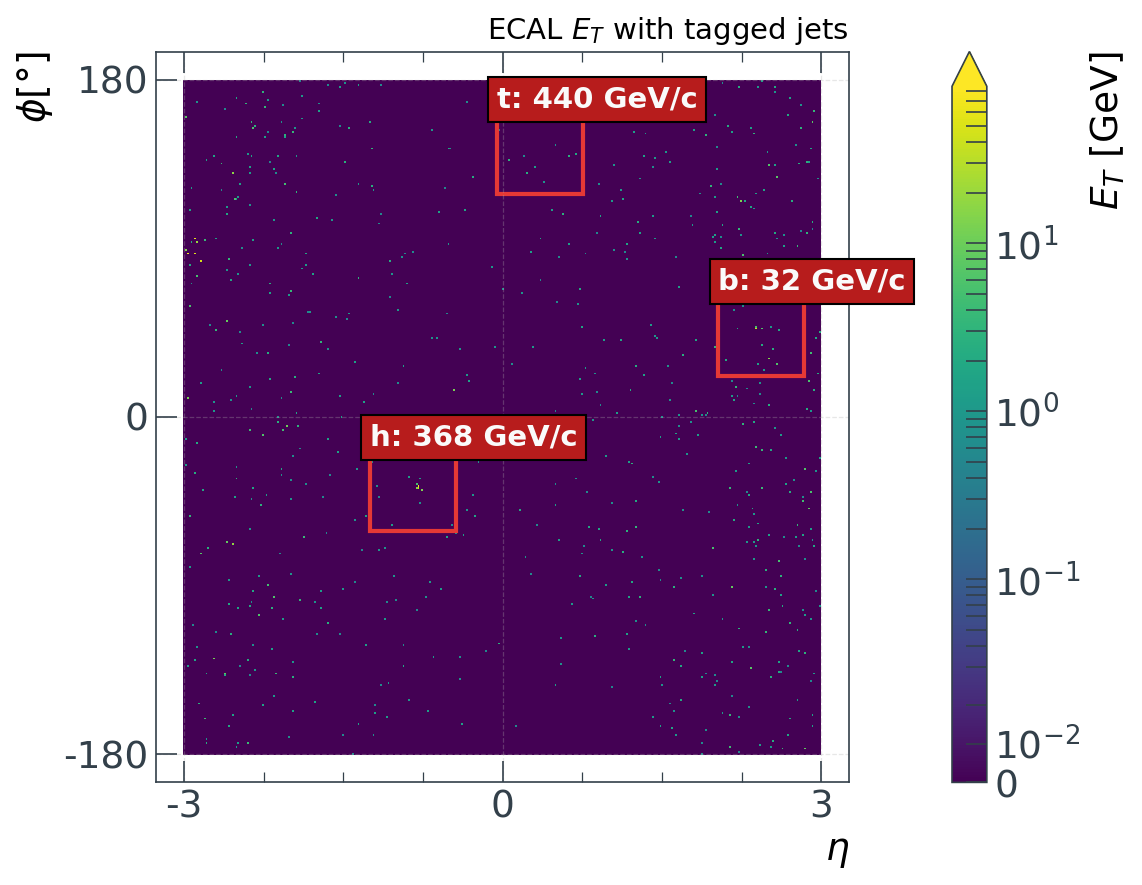}
  \includegraphics[width=.49\linewidth]{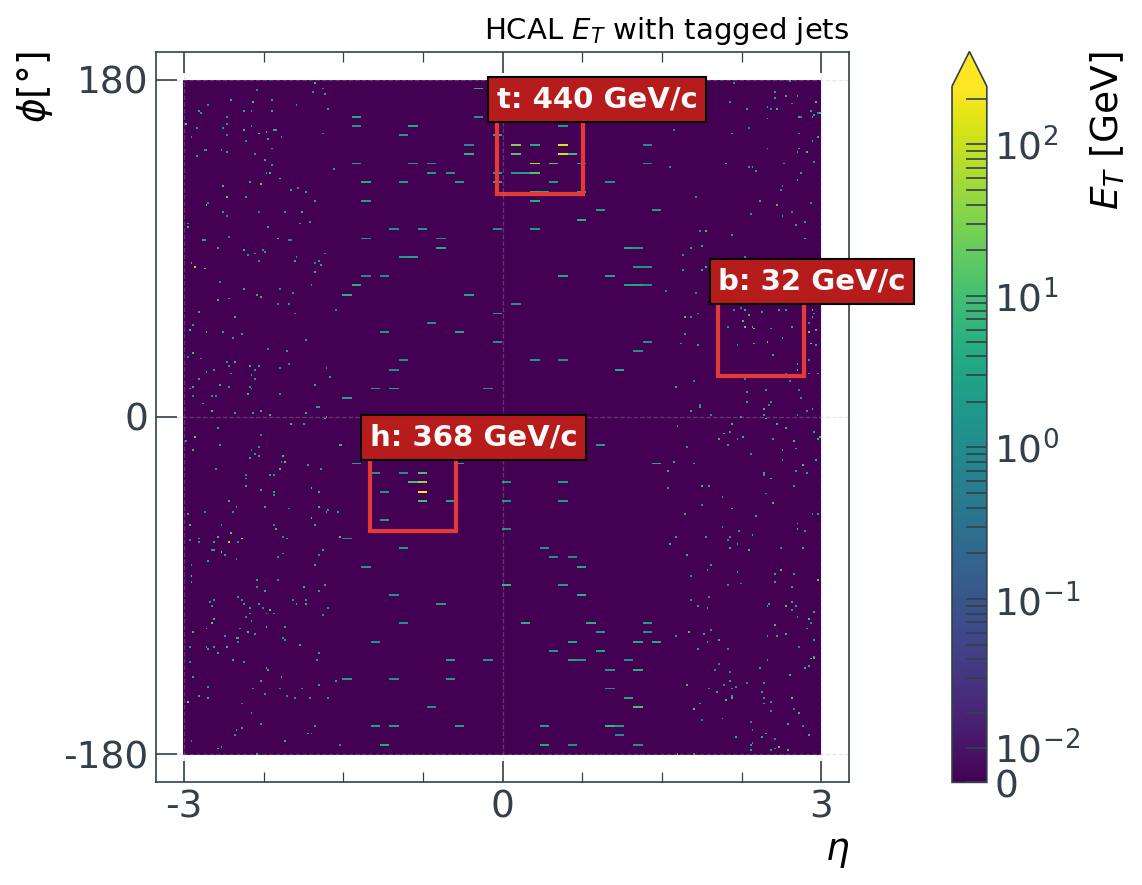}
  \caption{Energy deposits in CMS ECAL and HCAL translated to a two-dimensional image, an example input to the SSD network. The red bounding boxes correspond to ground truth with target label and momentum.}
  \label{fig:example}
\end{figure}

As a proof of concept, we investigate the tagging of the bottom (b) W boson (W), Higgs boson (H), or top quark (t) jet. An example input, energy deposits translated to two-dimensional images with two channels (corresponding to ECAL and HCAL) together with marked ground truth bounding boxes is shown in Figure~\ref{fig:example}.

\section{Model, implementation and training procedure}\label{SEC:Model}

\begin{figure}
  \centering
  \includegraphics[width=\linewidth]{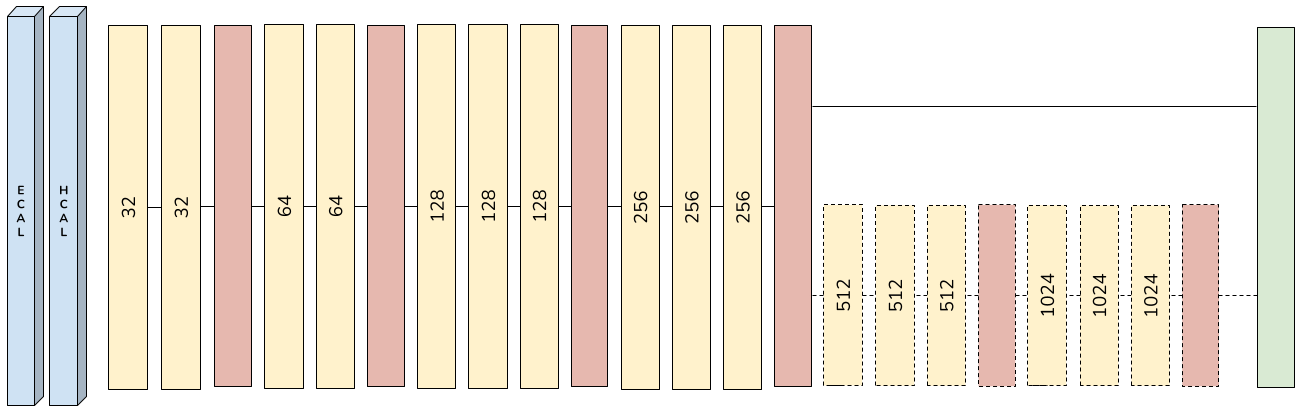}
  \caption{Jet-SSD architecture: input (in blue), convolution block, i.e. convolution layer followed by batch normalization and PReLU activation (in yellow), average pooling (in red) and output layer (in green). The numbers indicate the number of output channels in each block. The part of the network highlighted by dashed lines is used only during the training step.}
  \label{fig:jetssd}
\end{figure}

The Jet-SSD architecture is shown in Figure~\ref{fig:jetssd}. Several modifications are applied to the original architecture~\cite{liu2016ssd}. Due to target hardware constraints, all filters in convolution layers are of size $3\times3$ with no dilatation and all pooling layers have $2\times2$ filters. Each convolution block is followed by batch normalization~\cite{ioffe2015batch, sari2020does} and parametric rectified linear unit (PReLU) layers. To compress the model we use half of the channels of the VGG-16 in each layer. We also remove bias from all convolution layers. The extra layers proposed by the original paper do not contribute to accurate detection due to the size of jets and thus they are removed at the training. Retaining the deeper layers in the base network does not show improvements in the final detection results either, but they are critical during training due to additional signal during back-propagation. Hence, we only purge them at inference.

The Jet-SSD network is implemented on an NVidia Tesla GPU using PyTorch~\cite{NEURIPS2019_9015}. For training, we use stochastic gradient descent with an initial learning rate of $10^{-3}$ with momentum set to $0.9$ and weight regularization to $0.0005$. We train the network for $100$ epochs with a batch size of $25$, decreasing the learning rate by a factor of $2$ after $20,30,50,60,70,80$ and $90$ epochs. We use $90$k and $30$k samples for training and validation, respectively. The training is performed in mixed-precision to speed up computation and distributed across $3$ GPUs.

The full precision network (FPN) is trained from scratch using Xavier uniform initialization~\cite{pmlr-v9-glorot10a} (which helps with the sparsity of the input) as the pre-trained classification models on the real-world ImageNet~\cite{5206848} dataset have little relation to our calorimeter images. A common challenge when training models from scratch is the insufficient amount of training data which may lead to overfitting. However, it is not a problem in our case: the training dataset is large enough and, if overfitting occurred, we can go back and generate an even larger one. For TWN training we find out that pre-loading trained FPN weights greatly speeds up the process. And per-layer and per-channel scaling factor $\alpha$ improves the results.

The final detection layer returns a classification label (background, b, W/H or t jet) and three regression values. Two of them correspond to the centre of the jet, i.e. offset in $\eta$ and $\phi$ plane from the anchor. The last one is jet mass regression which is an example of an auxiliary function that Jet-SSD can be tasked with.

\begin{figure}
  \centering
  \includegraphics[width=\linewidth]{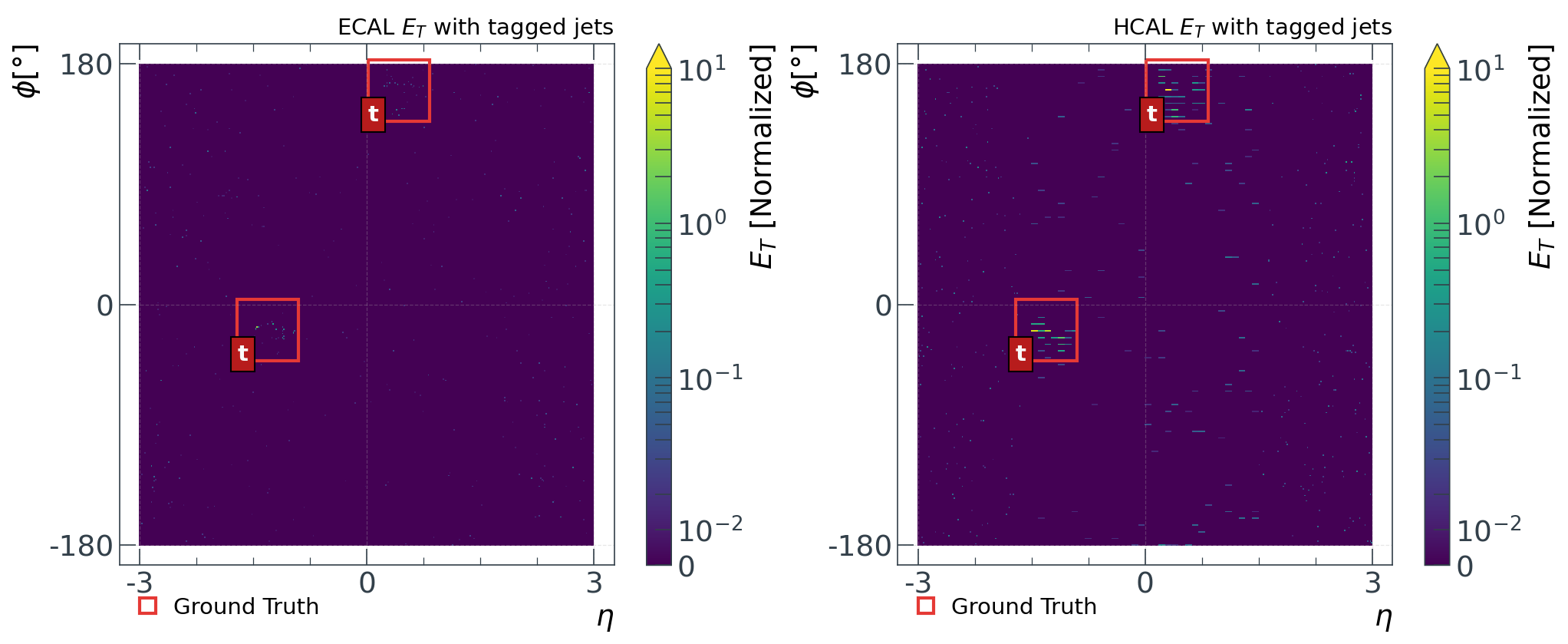}
  \includegraphics[width=\linewidth]{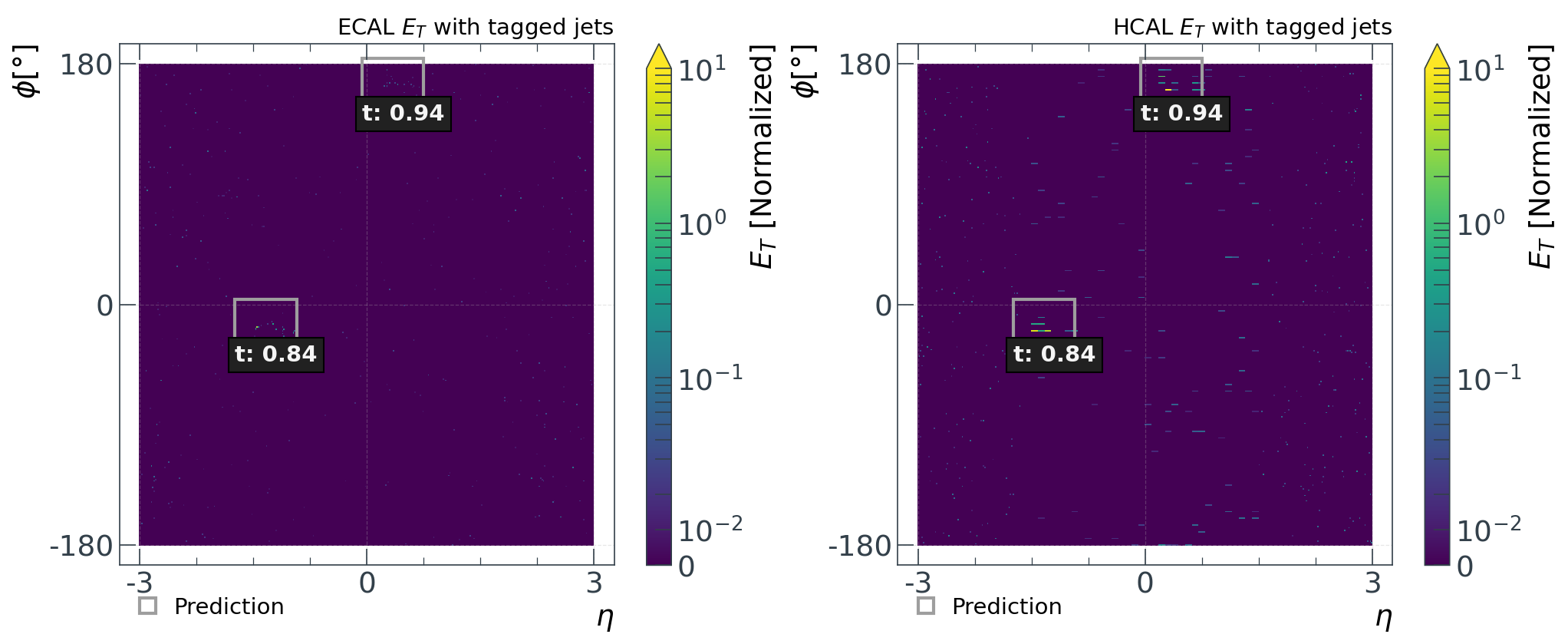}
  \caption{An example of the Jet-SSD at inference for one event with the input calorimeter image and highlighted true labels (top) and prediction bounding boxes (bottom). The fractional number next to the categorical label corresponds to the network confidence score.}
  \label{fig:inference}
\end{figure}

\begin{figure}
  \centering
  \includegraphics[width=.55\linewidth]{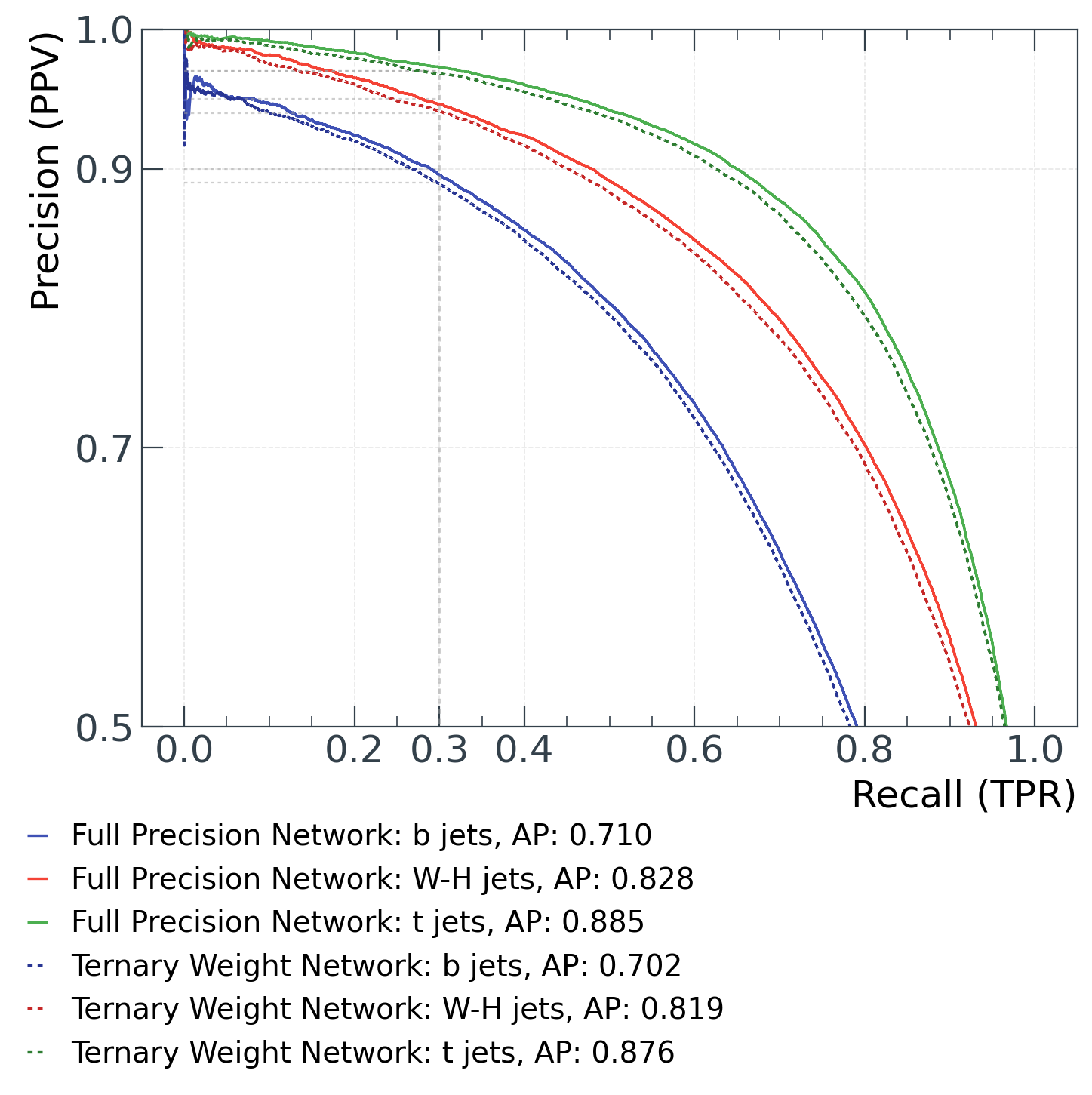}
  \caption{Jet-SSD PR curve and AP values for three target categories (b, W-H and t) and two weights precision (FPN and TWN).}
  \label{fig:pr}
\end{figure}

\section{Results}\label{SEC:Results}

An example of the Jet-SSD in action is shown in Figure~\ref{fig:inference}. Jet-SSD outputs predicted categorical label of the object, confidence and bounding boxes. In object detection true positive is defined as prediction with category equal to the ground truth label and Intersection over Union (IoU) above the predefined threshold, in our case $0.5$. Successful prediction meets both criteria, otherwise, it is considered as a false negative.

To evaluate the model we use precision and recall (true positive rate), and average precision (AP) metric, which is computed for each category separately. Classification tasks usually report on the receiver operator characteristic (ROC) curve, which is a function of the false positive rate (fall-out or the background efficiency) as a function of the true positive rate (sensitivity or signal efficiency). In the case of object detection, the false positive rate is not very informative as there is a big imbalance between positive and negative class (there are no objects in most locations). Thus, the false positive rate is replaced by precision or positive predictive value (PPV). Intuitively, precision measures how accurate the predictions while recall measures the quality of the positive predictions. To draw a precision-recall (PR) curve, the predictions are first sorted in order of confidence followed by calculation of positive predictive value and true positive rate for each confidence threshold. For the relationship between ROC and PR curve, see~\cite{rocpr}.

The PR curve of Jet-SSD, evaluated on a held-out test dataset consisting of $90$k samples, is shown in Figure~\ref{fig:pr}. The TWN results are closely matching the results of the FPN, which is reflected in an AP score. To calculate the value of AP, the maximum precision is calculated for the recall values that range from $0$ to $1$ with a step size of $0.1$ and finally averaging over the results. From the PR curve, we can conclude that t jets are the easiest to identify while b jets detection is lacking. The result is not surprising for two reasons. Firstly, b jets have a lower momentum threshold, making the energy deposits more challenging to detect. Secondly, CNN based object detection is more challenging as the scale of the target object decreases; and b jets have a smaller radius than t, W and H jets. The latter issue can be further mitigated as small scale object detection is an active research field in machine learning (for example~\cite{rabbi2020small}).

\begin{figure}
  \centering
  \includegraphics[width=.325\linewidth]{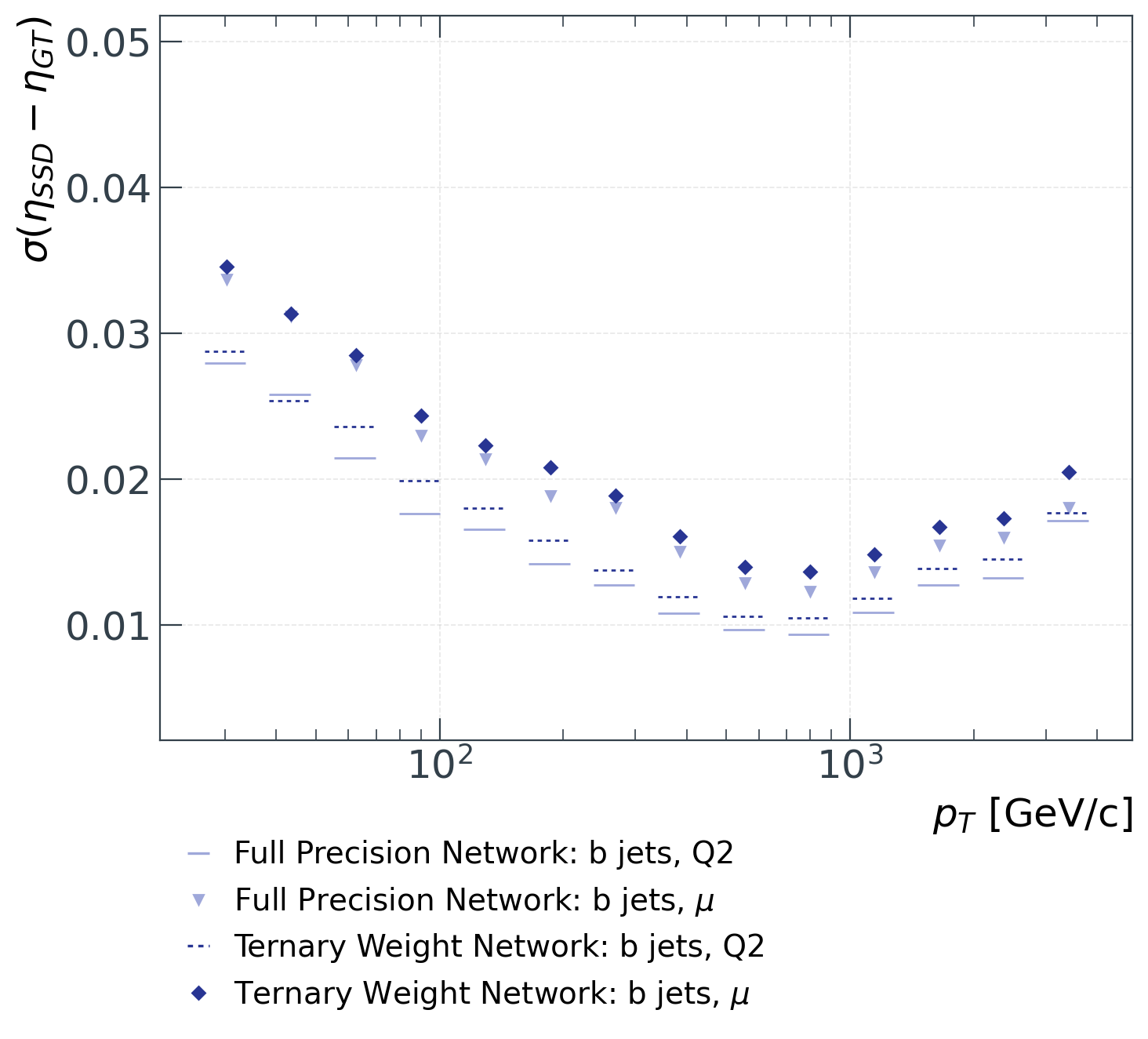}
  \includegraphics[width=.325\linewidth]{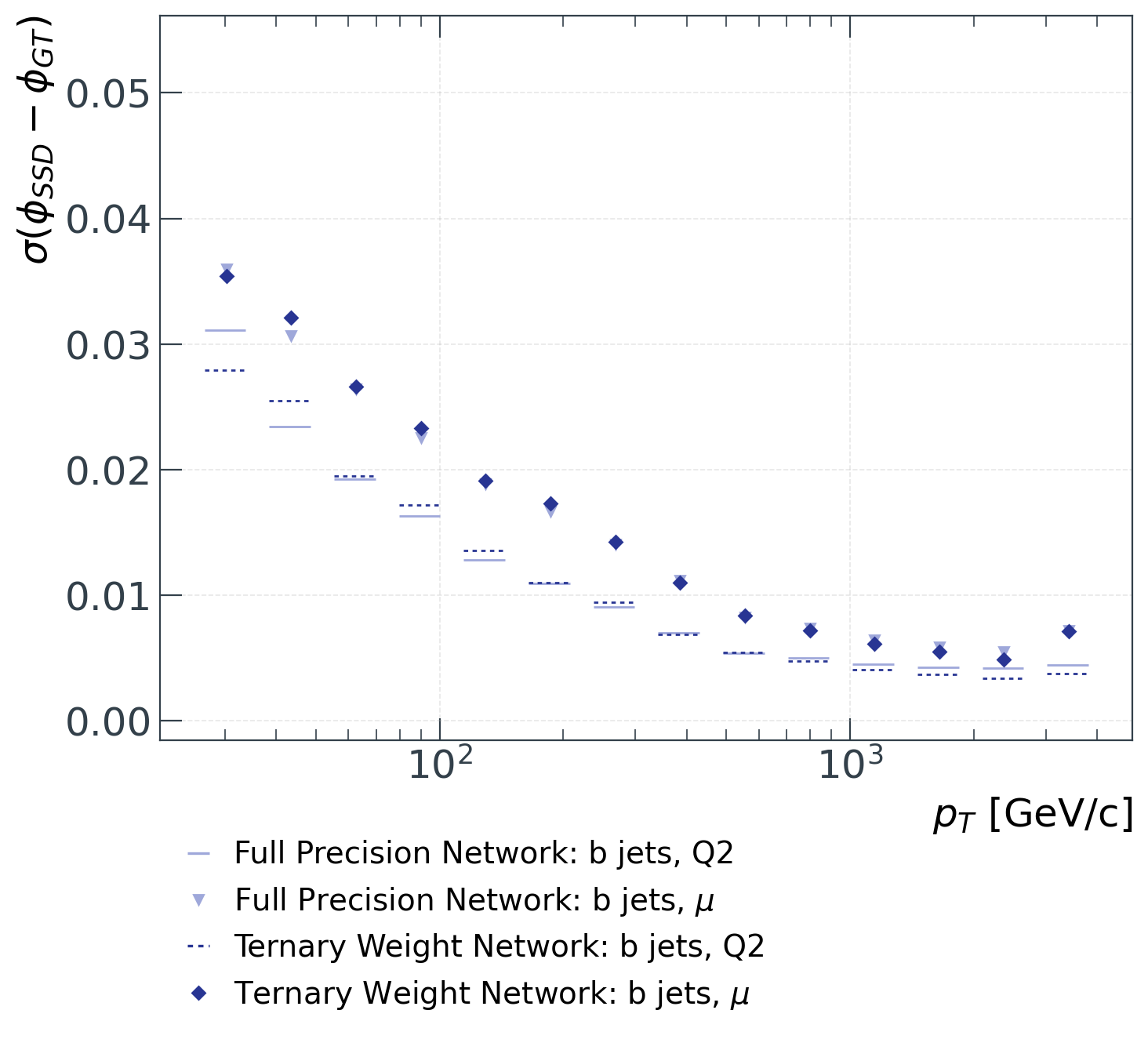}
  \includegraphics[width=.325\linewidth]{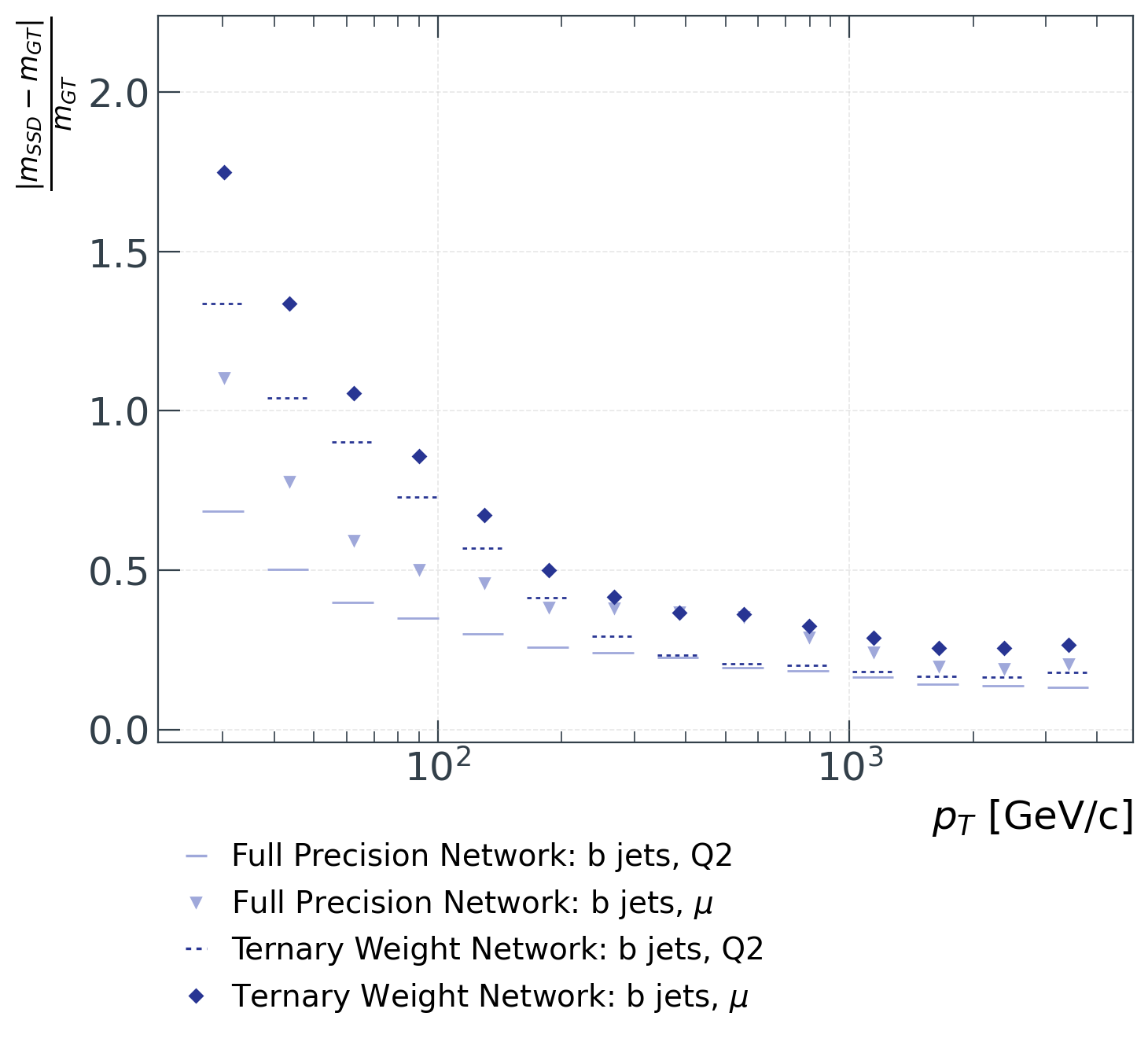}
  
  \includegraphics[width=.325\linewidth]{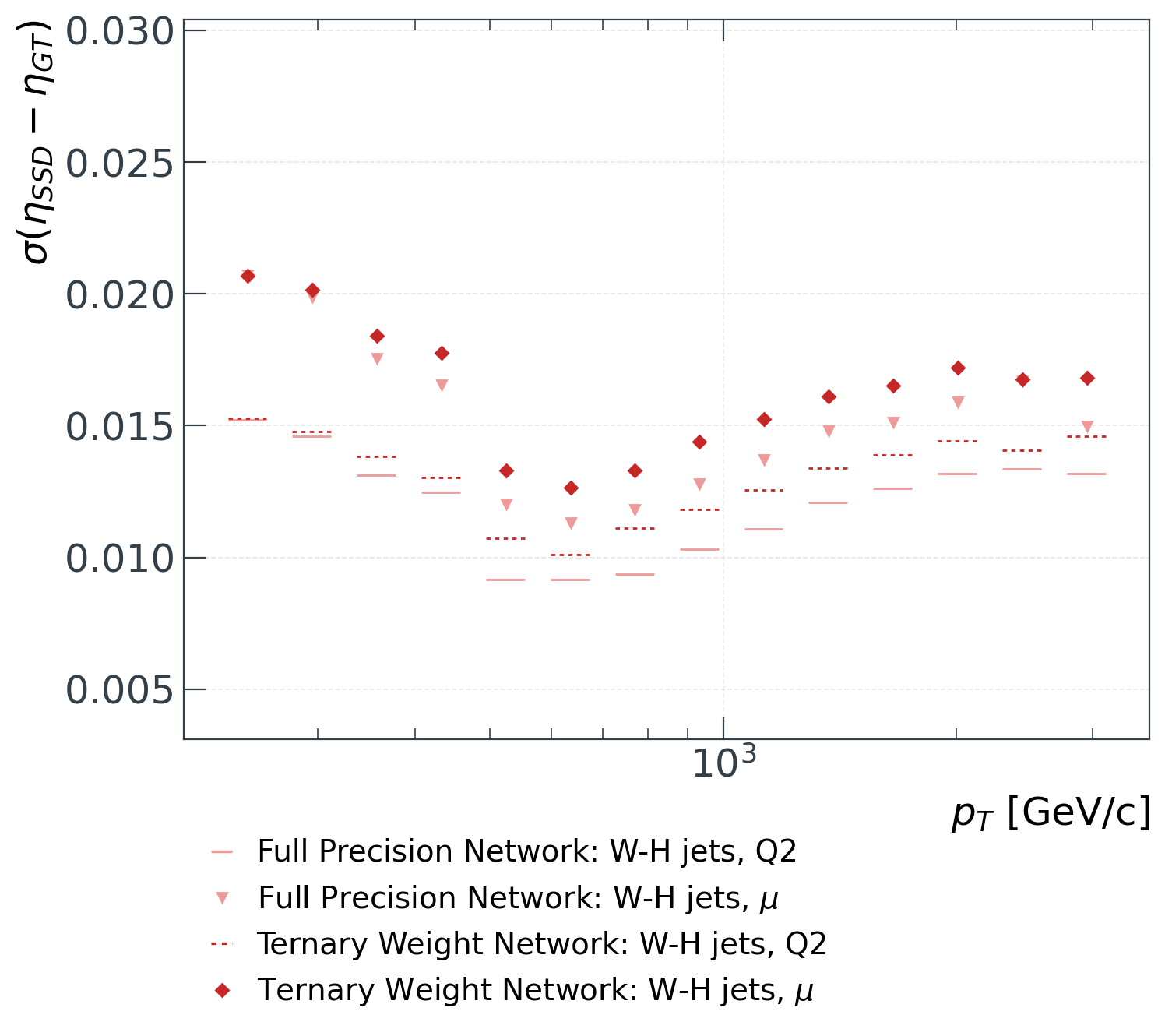}
  \includegraphics[width=.325\linewidth]{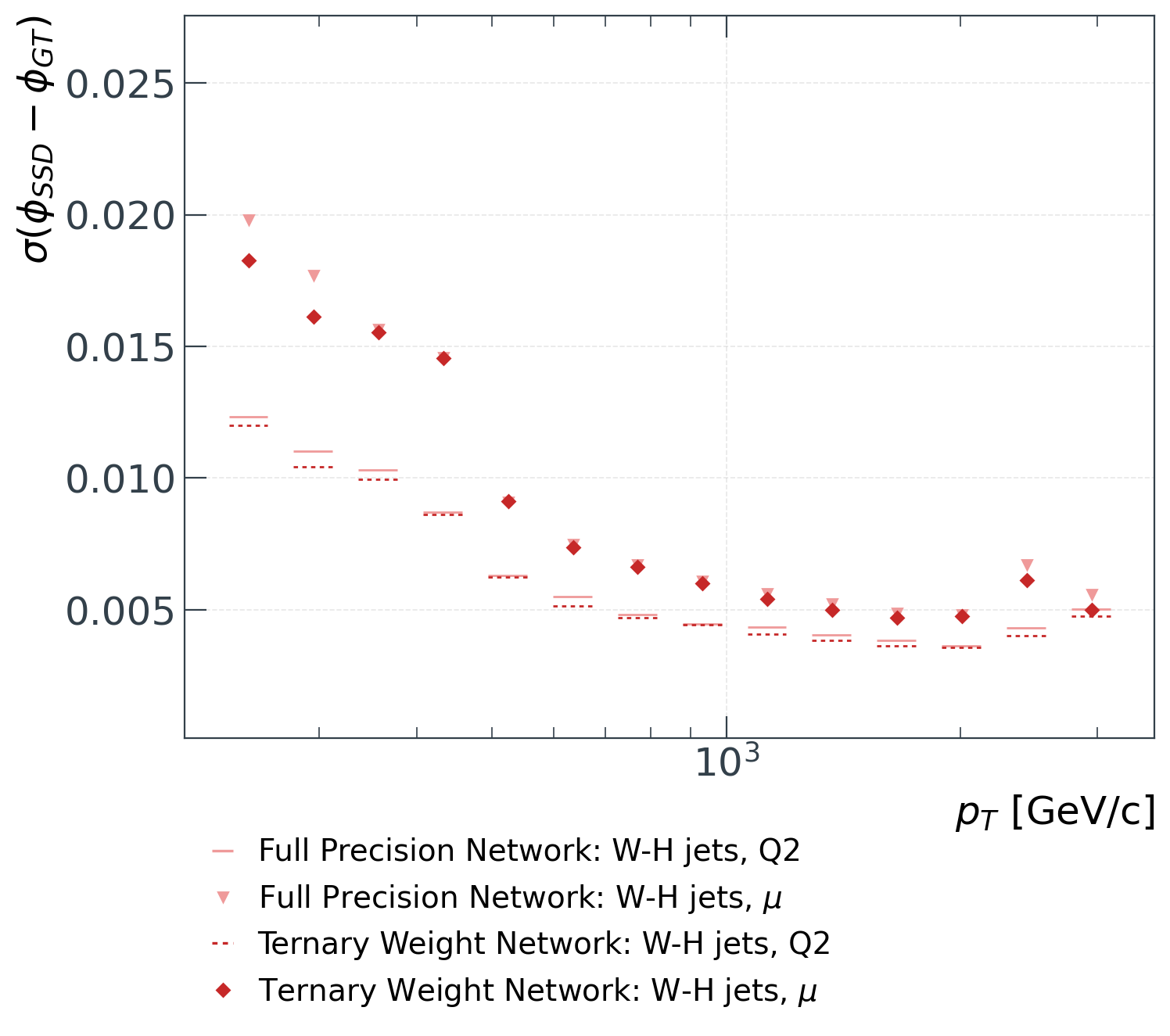}
  \includegraphics[width=.325\linewidth]{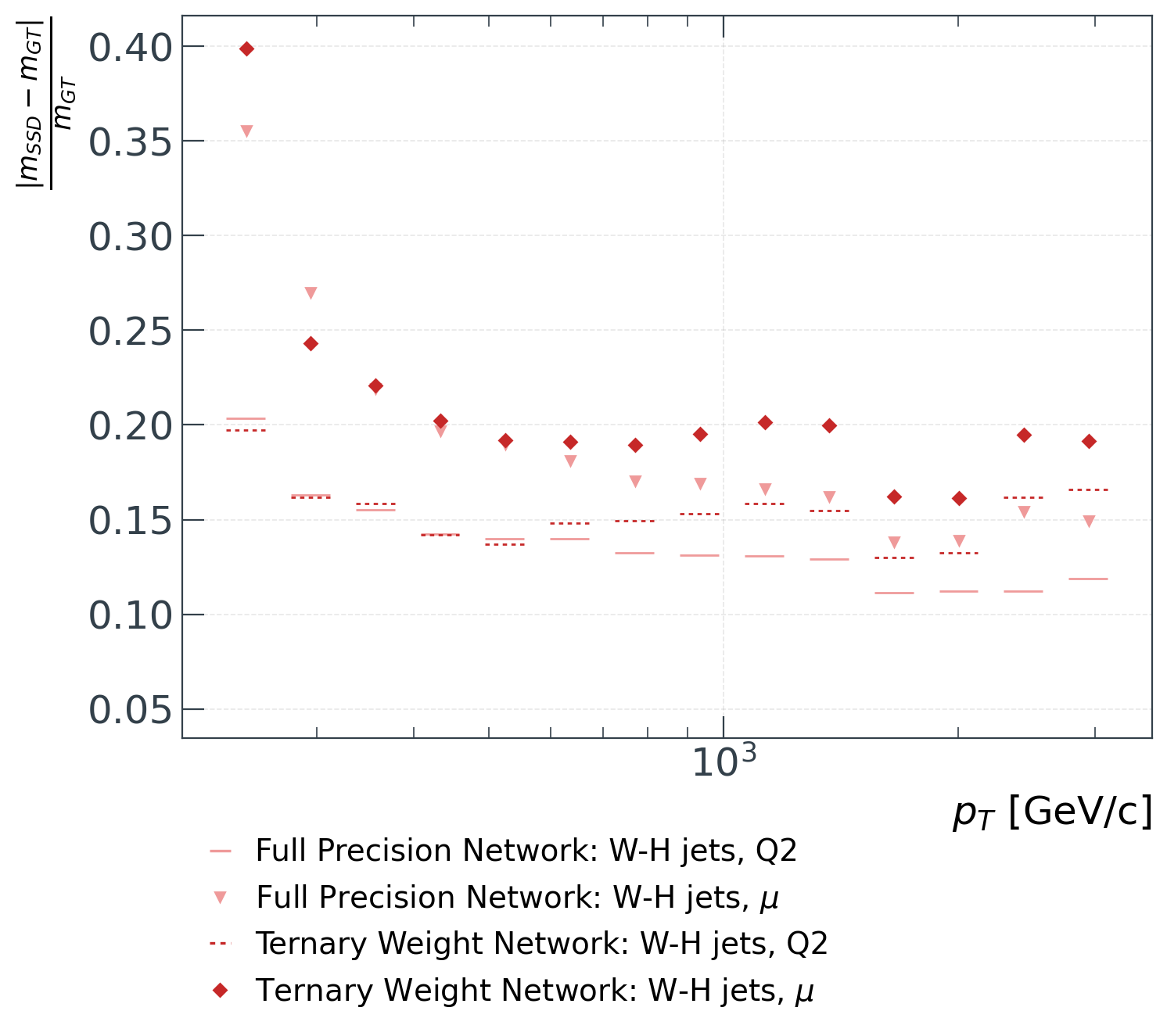}
  
  \includegraphics[width=.325\linewidth]{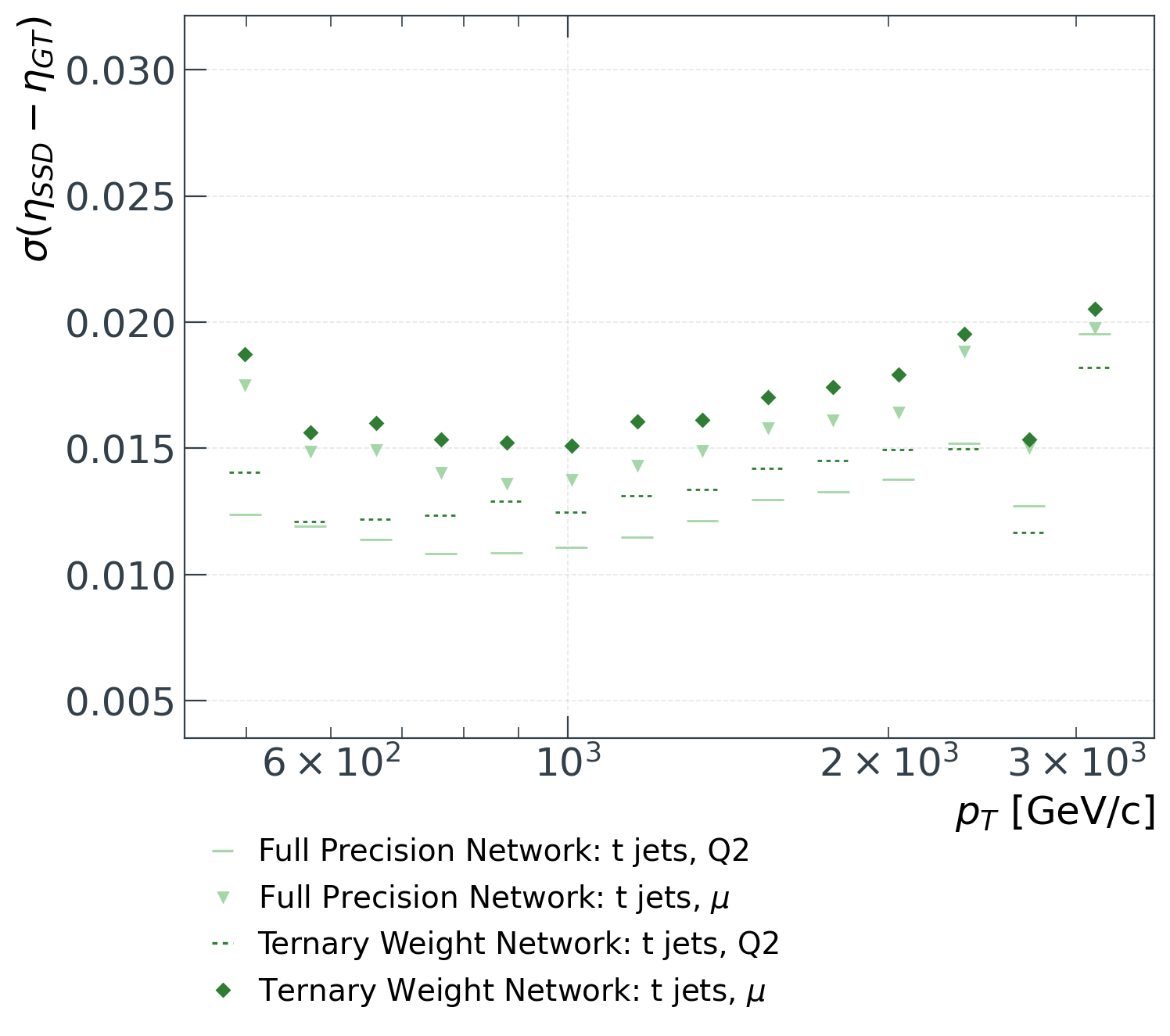}
  \includegraphics[width=.325\linewidth]{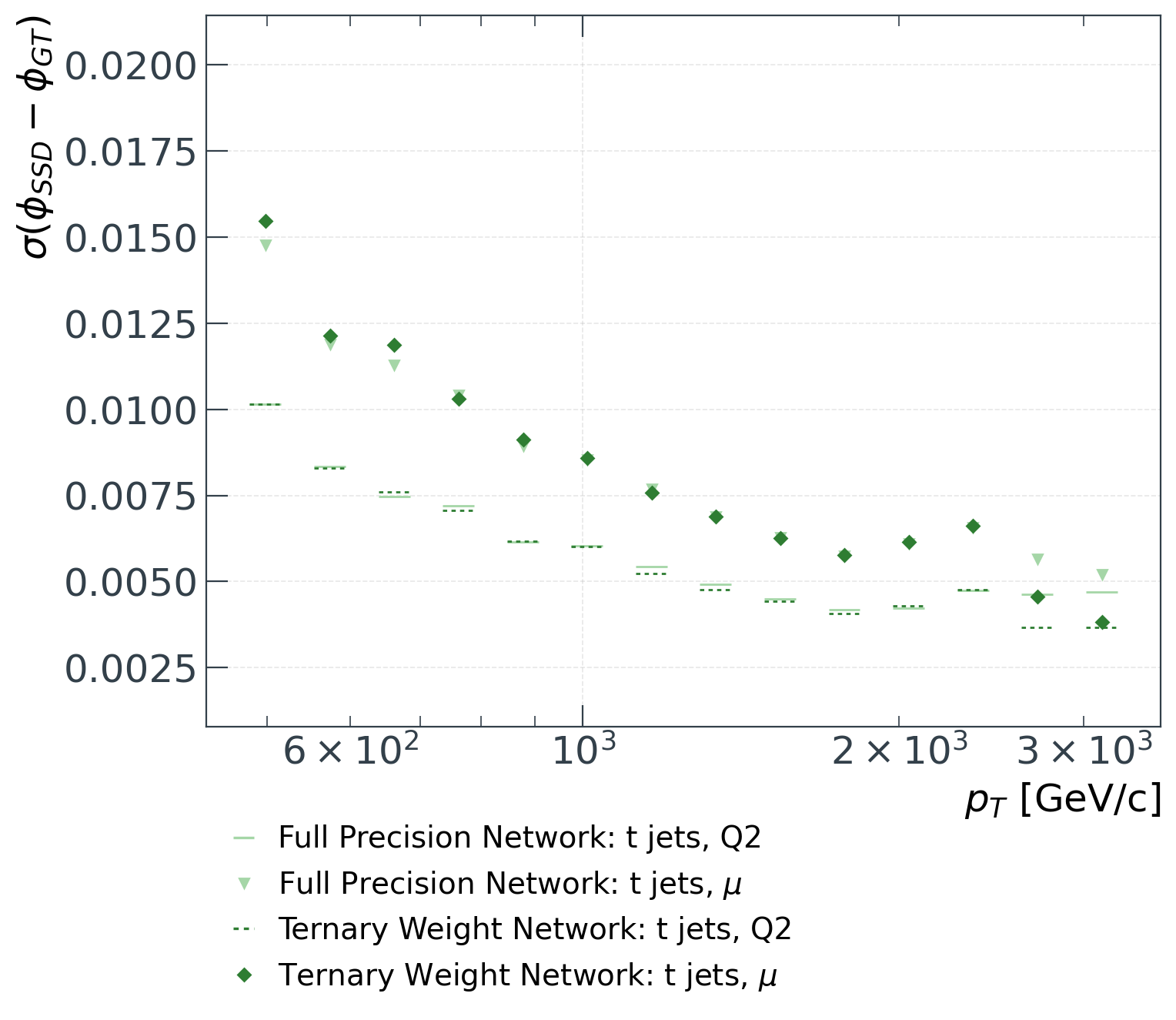}
  \includegraphics[width=.325\linewidth]{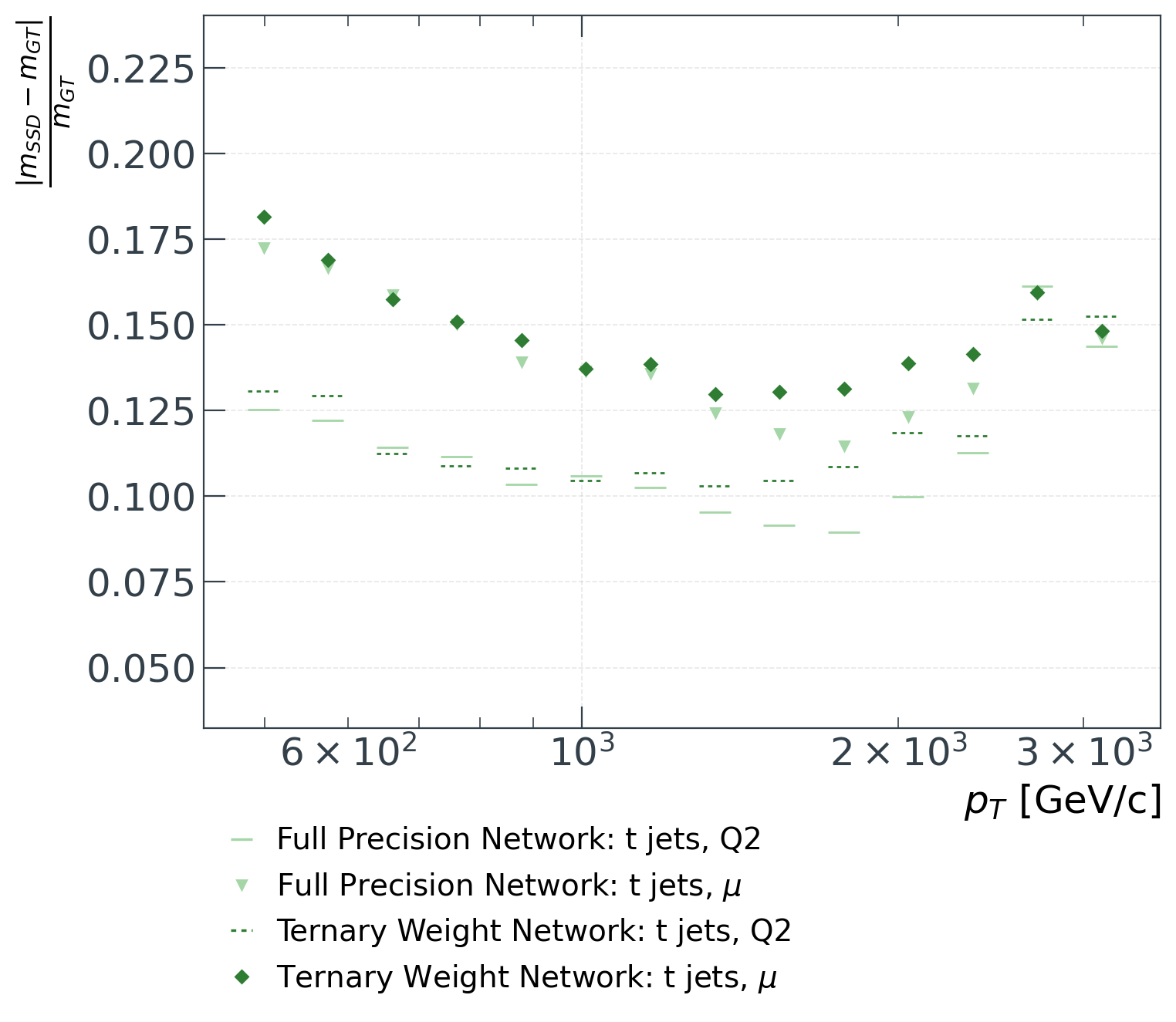}
  
  \caption{The mean ($\mu$) and median (Q2) localization error in $\eta$ (left column), $\phi$ (centre column) and relative error in mass regression between ground truth (GT) and Jet-SSD output (SSD) for FPN and TWN versions. The results are reported for each class independently: b jets (top row), W-H jets (centre row) and t jest (bottom row). All results are calculated as a function of p$_T$.}
  \label{fig:delta}
\end{figure}

Finally, we report the mean and median localization error in $\phi$ and $\eta$ and the relative error in mass regression. These results are shown in Figure~\ref{fig:delta}. The $\phi$ localization error is smaller than $\eta$ due to input information loss. Remind that we limit input in $\eta$ dimension. In the case when the jet centre is close to the edge, i.e. $|\eta| \approx 3.0$, part of the information is lost beyond image boundaries. Due to the cylindrical structure of the detector, this is not happening in $\phi$ dimension. Furthermore, we notice that the error does not decrease with p$_T$ for $\eta$ for which we don not find a reason. Finally, the mass regression relative error can be further decreased with re-balancing of the SSD training loss, i.e. increasing regression error contribution to back-propagation by introducing a new scaling hyper-parameter $\beta$: $loss = classification + localization + \beta \times auxiliary$, where $\beta > 1$.

\section{Conclusions}\label{SEC:Conclusions}
In this paper, we introduced Jet-SSD, a deep learning network able to simultaneously localize, tag and estimate the mass of jets, a collimated spray of particles produced in high energy physics experiments. We showed that the compressed model via quantized weights to ternary values with layer- and channel-dependent scaling factor closely matches the performance of the full precision model. We seek to examine the performance of the network on dedicated hardware.

\section*{Acknowledgments}\label{SEC:Acknowledgments}
A.~A.~P., M.~P., S.~S. and V.~L. are supported by the European Research Council (ERC) under the European Union's Horizon 2020 research and innovation program (grant agreement n$^o$ 772369). V.~L. is supported by Zenseact under the CERN Knowledge Transfer Group. A.~A.~P. is supported by CEVA under the CERN Knowledge Transfer Group. We thank Simons Foundation, Flatiron Institute and Ian Fisk for granting access to computing resources used for this project.

\bibliography{reference}

\begin{thebibliography}{83}

\bibitem{butterworth2008jet}
J.M. Butterworth, A.R. Davison, M.~Rubin, G.P. Salam, Physical review letters
  \textbf{100}, 242001 (2008)

\bibitem{skiba2007using}
W.~Skiba, D.~Tucker-Smith, Physical Review D \textbf{75}, 115010 (2007)

\bibitem{khachatryan2014search}
V.~Khachatryan, A.M. Sirunyan, A.~Tumasyan, W.~Adam, T.~Bergauer,
  M.~Dragicevic, J.~Er{\"o}, C.~Fabjan, M.~Friedl, R.~Fruehwirth et~al.,
  Journal of High Energy Physics \textbf{2014}, 173 (2014)

\bibitem{aad2015search}
G.~Aad, B.~Abbott, J.~Abdallah, R.~Aben, M.~Abolins, O.~AbouZeid,
  H.~Abramowicz, H.~Abreu, R.~Abreu, Y.~Abulaiti et~al., Journal of High Energy
  Physics \textbf{2015}, 1 (2015)

\bibitem{adams2015towards}
D.~Adams, A.~Arce, L.~Asquith, M.~Backovic, T.~Barillari, P.~Berta,
  D.~Bertolini, A.~Buckley, J.~Butterworth, R.C. Toro et~al., The European
  Physical Journal C \textbf{75}, 1 (2015)

\bibitem{abdesselam2011boosted}
A.~Abdesselam, A.~Belyaev, E.B. Kuutmann, U.~Bitenc, G.~Brooijmans,
  J.~Butterworth, P.B. de~Renstrom, D.B. Franzosi, R.~Buckingham, B.~Chapleau
  et~al., The European Physical Journal C \textbf{71}, 1 (2011)

\bibitem{altheimer2012jet}
A.~Altheimer, S.~Arora, L.~Asquith, G.~Brooijmans, J.~Butterworth,
  M.~Campanelli, B.~Chapleau, A.~Cholakian, J.~Chou, M.~Dasgupta et~al.,
  Journal of Physics G: Nuclear and Particle Physics \textbf{39}, 063001 (2012)

\bibitem{altheimer2014boosted}
A.~Altheimer, A.~Arce, L.~Asquith, J.B. Mayes, E.B. Kuutmann, J.~Berger,
  D.~Bjergaard, L.~Bryngemark, A.~Buckley, J.~Butterworth et~al., The European
  Physical Journal C \textbf{74}, 1 (2014)

\bibitem{plehn2010stop}
T.~Plehn, M.~Spannowsky, M.~Takeuchi, D.~Zerwas, Journal of High Energy Physics
  \textbf{2010}, 1 (2010)

\bibitem{larkoski2014soft}
A.J. Larkoski, S.~Marzani, G.~Soyez, J.~Thaler, Journal of High Energy Physics
  \textbf{2014}, 146 (2014)

\bibitem{thaler2011identifying}
J.~Thaler, K.~Van~Tilburg, Journal of High Energy Physics \textbf{2011}, 15
  (2011)

\bibitem{larkoski2013energy}
A.J. Larkoski, G.P. Salam, J.~Thaler, Journal of High Energy Physics
  \textbf{2013}, 108 (2013)

\bibitem{krohn2010jet}
D.~Krohn, J.~Thaler, L.T. Wang, Journal of High Energy Physics \textbf{2010},
  84 (2010)

\bibitem{ellis2010recombination}
S.D. Ellis, C.K. Vermilion, J.R. Walsh, Physical Review D \textbf{81}, 094023
  (2010)

\bibitem{dasgupta2013towards}
M.~Dasgupta, A.~Fregoso, S.~Marzani, G.P. Salam, Journal of High Energy Physics
  \textbf{2013}, 29 (2013)

\bibitem{dasgupta2013jet}
M.~Dasgupta, A.~Fregoso, S.~Marzani, A.~Powling, The European Physical Journal
  C \textbf{73}, 1 (2013)

\bibitem{dasgupta2015jet}
M.~Dasgupta, A.~Powling, A.~Siodmok, Journal of High Energy Physics
  \textbf{2015}, 1 (2015)

\bibitem{cogan2015jet}
J.~Cogan, M.~Kagan, E.~Strauss, A.~Schwarztman, Journal of High Energy Physics
  \textbf{2015}, 118 (2015)

\bibitem{pearkes2017jet}
J.~Pearkes, W.~Fedorko, A.~Lister, C.~Gay, arXiv preprint arXiv:1704.02124
  (2017)

\bibitem{baldi2016jet}
P.~Baldi, K.~Bauer, C.~Eng, P.~Sadowski, D.~Whiteson, Physical Review D
  \textbf{93}, 094034 (2016)

\bibitem{macaluso2018pulling}
S.~Macaluso, D.~Shih, Journal of High Energy Physics \textbf{2018}, 1 (2018)

\bibitem{almeida2015playing}
L.G. Almeida, M.~Backovi{\'c}, M.~Cliche, S.J. Lee, M.~Perelstein, Journal of
  High Energy Physics \textbf{2015}, 1 (2015)

\bibitem{de2016jet}
L.~de~Oliveira, M.~Kagan, L.~Mackey, B.~Nachman, A.~Schwartzman, Journal of
  High Energy Physics \textbf{2016}, 1 (2016)

\bibitem{guest2016jet}
D.~Guest, J.~Collado, P.~Baldi, S.C. Hsu, G.~Urban, D.~Whiteson, Physical
  Review D \textbf{94}, 112002 (2016)

\bibitem{barnard2017parton}
J.~Barnard, E.N. Dawe, M.J. Dolan, N.~Rajcic, Physical Review D \textbf{95},
  014018 (2017)

\bibitem{butter2018deep}
A.~Butter, G.~Kasieczka, T.~Plehn, M.~Russell, SciPost Phys \textbf{5}, 028
  (2018)

\bibitem{komiske2017deep}
P.T. Komiske, E.M. Metodiev, M.D. Schwartz, Journal of High Energy Physics
  \textbf{2017}, 110 (2017)

\bibitem{lin2018boosting}
J.~Lin, M.~Freytsis, I.~Moult, B.~Nachman, Journal of High Energy Physics
  \textbf{2018}, 1 (2018)

\bibitem{kasieczka2019machine}
G.~Kasieczka, T.~Plehn, A.~Butter, K.~Cranmer, D.~Debnath, B.M. Dillon,
  M.~Fairbairn, D.A. Faroughy, W.~Fedorko, C.~Gay et~al., arXiv preprint
  arXiv:1902.09914  (2019)

\bibitem{kasieczka2017deep}
G.~Kasieczka, T.~Plehn, M.~Russell, T.~Schell, Journal of High Energy Physics
  \textbf{2017}, 6 (2017)

\bibitem{lecun1998gradient}
Y.~LeCun, L.~Bottou, Y.~Bengio, P.~Haffner, Proceedings of the IEEE
  \textbf{86}, 2278 (1998)

\bibitem{cacciari2012fastjet}
M.~Cacciari, G.P. Salam, G.~Soyez, The European Physical Journal C \textbf{72},
  1 (2012)

\bibitem{Sirunyan:2019wwa}
A.M. Sirunyan et~al. (CMS), Comput. Softw. Big Sci. \textbf{4}, 10 (2020),
  \texttt{1912.06046}

\bibitem{sermanet2013overfeat}
P.~Sermanet, D.~Eigen, X.~Zhang, M.~Mathieu, R.~Fergus, Y.~LeCun, arXiv
  preprint arXiv:1312.6229  (2013)

\bibitem{zhang2016joint}
K.~Zhang, Z.~Zhang, Z.~Li, Y.~Qiao, IEEE Signal Processing Letters \textbf{23},
  1499 (2016)

\bibitem{zhang2016faster}
L.~Zhang, L.~Lin, X.~Liang, K.~He, \emph{Is faster R-CNN doing well for
  pedestrian detection?}, in \emph{European conference on computer vision}
  (Springer, 2016), pp. 443--457

\bibitem{zou2019object}
Z.~Zou, Z.~Shi, Y.~Guo, J.~Ye, arXiv preprint arXiv:1905.05055  (2019)

\bibitem{liu2020deep}
L.~Liu, W.~Ouyang, X.~Wang, P.~Fieguth, J.~Chen, X.~Liu, M.~Pietik{\"a}inen,
  International journal of computer vision \textbf{128}, 261 (2020)

\bibitem{redmon2017yolo9000}
J.~Redmon, A.~Farhadi, \emph{YOLO9000: better, faster, stronger}, in
  \emph{Proceedings of the IEEE conference on computer vision and pattern
  recognition} (2017), pp. 7263--7271

\bibitem{redmon2016you}
J.~Redmon, S.~Divvala, R.~Girshick, A.~Farhadi, \emph{You only look once:
  Unified, real-time object detection}, in \emph{Proceedings of the IEEE
  conference on computer vision and pattern recognition} (2016), pp. 779--788

\bibitem{fu2017dssd}
C.Y. Fu, W.~Liu, A.~Ranga, A.~Tyagi, A.C. Berg, arXiv preprint arXiv:1701.06659
   (2017)

\bibitem{zhou2019objects}
X.~Zhou, D.~Wang, P.~Kr{\"a}henb{\"u}hl, arXiv preprint arXiv:1904.07850
  (2019)

\bibitem{lin2017focal}
T.Y. Lin, P.~Goyal, R.~Girshick, K.~He, P.~Doll{\'a}r, \emph{Focal loss for
  dense object detection}, in \emph{Proceedings of the IEEE international
  conference on computer vision} (2017), pp. 2980--2988

\bibitem{girshick2014rich}
R.~Girshick, J.~Donahue, T.~Darrell, J.~Malik, \emph{Rich feature hierarchies
  for accurate object detection and semantic segmentation}, in
  \emph{Proceedings of the IEEE conference on computer vision and pattern
  recognition} (2014), pp. 580--587

\bibitem{ren2015faster}
S.~Ren, K.~He, R.~Girshick, J.~Sun, arXiv preprint arXiv:1506.01497  (2015)

\bibitem{girshick2015fast}
R.~Girshick, \emph{Fast r-cnn}, in \emph{Proceedings of the IEEE international
  conference on computer vision} (2015), pp. 1440--1448

\bibitem{dai2016r}
J.~Dai, Y.~Li, K.~He, J.~Sun, arXiv preprint arXiv:1605.06409  (2016)

\bibitem{xu2018deep}
H.~Xu, X.~Lv, X.~Wang, Z.~Ren, N.~Bodla, R.~Chellappa, \emph{Deep regionlets
  for object detection}, in \emph{Proceedings of the European Conference on
  Computer Vision (ECCV)} (2018), pp. 798--814

\bibitem{liu2016ssd}
W.~Liu, D.~Anguelov, D.~Erhan, C.~Szegedy, S.~Reed, C.Y. Fu, A.C. Berg,
  \emph{Ssd: Single shot multibox detector}, in \emph{European conference on
  computer vision} (Springer, 2016), pp. 21--37

\bibitem{simonyan2014very}
K.~Simonyan, A.~Zisserman, arXiv preprint arXiv:1409.1556  (2014)

\bibitem{cheng2017survey}
Y.~Cheng, D.~Wang, P.~Zhou, T.~Zhang, arXiv preprint arXiv:1710.09282  (2017)

\bibitem{lecun1989optimal}
Y.~LeCun, J.S. Denker, S.A. Solla, R.E. Howard, L.D. Jackel, \emph{Optimal
  brain damage.}, in \emph{NIPs} (Citeseer, 1989), Vol.~2, pp. 598--605

\bibitem{han2015deep}
S.~Han, H.~Mao, W.J. Dally, arXiv preprint arXiv:1510.00149  (2015)

\bibitem{louizos2017learning}
C.~Louizos, M.~Welling, D.P. Kingma, arXiv preprint arXiv:1712.01312  (2017)

\bibitem{sironi2014learning}
A.~Sironi, B.~Tekin, R.~Rigamonti, V.~Lepetit, P.~Fua, IEEE transactions on
  pattern analysis and machine intelligence \textbf{37}, 94 (2014)

\bibitem{denton2014exploiting}
E.~Denton, W.~Zaremba, J.~Bruna, Y.~LeCun, R.~Fergus, arXiv preprint
  arXiv:1404.0736  (2014)

\bibitem{jaderberg2014speeding}
M.~Jaderberg, A.~Vedaldi, A.~Zisserman, arXiv preprint arXiv:1405.3866  (2014)

\bibitem{szegedy2015going}
C.~Szegedy, W.~Liu, Y.~Jia, P.~Sermanet, S.~Reed, D.~Anguelov, D.~Erhan,
  V.~Vanhoucke, A.~Rabinovich, \emph{Going deeper with convolutions}, in
  \emph{Proceedings of the IEEE conference on computer vision and pattern
  recognition} (2015), pp. 1--9

\bibitem{howard2017mobilenets}
A.G. Howard, M.~Zhu, B.~Chen, D.~Kalenichenko, W.~Wang, T.~Weyand,
  M.~Andreetto, H.~Adam, arXiv preprint arXiv:1704.04861  (2017)

\bibitem{iandola2016squeezenet}
F.N. Iandola, S.~Han, M.W. Moskewicz, K.~Ashraf, W.J. Dally, K.~Keutzer, arXiv
  preprint arXiv:1602.07360  (2016)

\bibitem{cohen2016group}
T.~Cohen, M.~Welling, \emph{Group equivariant convolutional networks}, in
  \emph{International conference on machine learning} (PMLR, 2016), pp.
  2990--2999

\bibitem{bucilua2006model}
C.~Buciluǎ, R.~Caruana, A.~Niculescu-Mizil, \emph{Model compression}, in
  \emph{Proceedings of the 12th ACM SIGKDD international conference on
  Knowledge discovery and data mining} (2006), pp. 535--541

\bibitem{courbariaux2015binaryconnect}
M.~Courbariaux, Y.~Bengio, J.P. David, arXiv preprint arXiv:1511.00363  (2015)

\bibitem{courbariaux2016binarized}
M.~Courbariaux, I.~Hubara, D.~Soudry, R.~El-Yaniv, Y.~Bengio, arXiv preprint
  arXiv:1602.02830  (2016)

\bibitem{zhou2016dorefa}
S.~Zhou, Y.~Wu, Z.~Ni, X.~Zhou, H.~Wen, Y.~Zou, arXiv preprint arXiv:1606.06160
   (2016)

\bibitem{rastegari2016xnor}
M.~Rastegari, V.~Ordonez, J.~Redmon, A.~Farhadi, \emph{Xnor-net: Imagenet
  classification using binary convolutional neural networks}, in \emph{European
  conference on computer vision} (Springer, 2016), pp. 525--542

\bibitem{hubara2017quantized}
I.~Hubara, M.~Courbariaux, D.~Soudry, R.~El-Yaniv, Y.~Bengio, The Journal of
  Machine Learning Research \textbf{18}, 6869 (2017)

\bibitem{li2016ternary}
F.~Li, B.~Zhang, B.~Liu, arXiv preprint arXiv:1605.04711  (2016)

\bibitem{zhu2016trained}
C.~Zhu, S.~Han, H.~Mao, W.J. Dally, arXiv preprint arXiv:1612.01064  (2016)

\bibitem{lee2017lognet}
E.H. Lee, D.~Miyashita, E.~Chai, B.~Murmann, S.S. Wong, \emph{Lognet:
  Energy-efficient neural networks using logarithmic computation}, in
  \emph{2017 IEEE International Conference on Acoustics, Speech and Signal
  Processing (ICASSP)} (IEEE, 2017), pp. 5900--5904

\bibitem{cai2017deep}
Z.~Cai, X.~He, J.~Sun, N.~Vasconcelos, \emph{Deep learning with low precision
  by half-wave gaussian quantization}, in \emph{Proceedings of the IEEE
  conference on computer vision and pattern recognition} (2017), pp. 5918--5926

\bibitem{cms2016cms}
{CMS Collaboration}, arXiv preprint arXiv:1609.02366  (2016)

\bibitem{bhimji2018deep}
W.~Bhimji, S.A. Farrell, T.~Kurth, M.~Paganini, E.~Racah et~al., \textbf{1085},
  042034 (2018)

\bibitem{collaboration2008cms}
{CMS Collaboration}, JInst \textbf{3}, S08004 (2008)

\bibitem{sjostrand2008brief}
T.~Sj{\"o}strand, S.~Mrenna, P.~Skands, Computer Physics Communications
  \textbf{178}, 852 (2008)

\bibitem{de2014delphes}
J.~De~Favereau, C.~Delaere, P.~Demin, A.~Giammanco, V.~Lemaitre, A.~Mertens,
  M.~Selvaggi, D.. Collaboration et~al., Journal of High Energy Physics
  \textbf{2014}, 57 (2014)

\bibitem{ioffe2015batch}
S.~Ioffe, C.~Szegedy, \emph{Batch normalization: Accelerating deep network
  training by reducing internal covariate shift} (2015), \texttt{1502.03167}

\bibitem{sari2020does}
E.~Sari, M.~Belbahri, V.P. Nia, \emph{How does batch normalization help binary
  training?} (2020), \texttt{1909.09139}

\bibitem{NEURIPS2019_9015}
A.~Paszke, S.~Gross, F.~Massa, A.~Lerer, J.~Bradbury, G.~Chanan, T.~Killeen,
  Z.~Lin, N.~Gimelshein, L.~Antiga et~al., \emph{{PyTorch: An Imperative Style,
  High-Performance Deep Learning Library}} (2019)

\bibitem{pmlr-v9-glorot10a}
X.~Glorot, Y.~Bengio, \emph{Understanding the difficulty of training deep
  feedforward neural networks}, in \emph{Proceedings of the thirteenth
  international conference on artificial intelligence and statistics} (JMLR
  Workshop and Conference Proceedings, 2010), pp. 249--256

\bibitem{5206848}
J.~{Deng}, W.~{Dong}, R.~{Socher}, L.~{Li}, {Kai Li}, {Li Fei-Fei},
  \emph{ImageNet: A large-scale hierarchical image database}, in \emph{2009
  IEEE Conference on Computer Vision and Pattern Recognition} (2009), pp.
  248--255

\bibitem{rocpr}
J.~Davis, M.~Goadrich, \emph{The relationship between Precision-Recall and ROC
  curves}, in \emph{Proceedings of the 23rd international conference on Machine
  learning} (2006), pp. 233--240

\bibitem{rabbi2020small}
J.~Rabbi, N.~Ray, M.~Schubert, S.~Chowdhury, D.~Chao, Remote Sensing
  \textbf{12}, 1432 (2020)

\end{thebibliography}

\end{document}